\documentclass[a4paper,11pt]{article}
\usepackage{pos}

\title{Learning about the QCD medium using electromagnetic and weak probes}

\author*[a]{Gojko Vujanovic}

\affiliation[a]{Wayne State University, 666 W Hancock St, Detroit, Michigan 48201, USA}

\emailAdd{vujanovic@wayne.edu}

\abstract{Recent theoretical developments concerning radiation of electromagnetic and weak probes in ultra-relativistic heavy-ion collisions are reviewed. These proceedings focus on electromagnetic probes and briefly cover weak probes. An outlook regarding the future use of electromagnetic probes is formulated whereby a quantitative Bayesian comparison, simultaneously employing electromagnetic and hadronic calculations of experimental observables against data, is highlighted as a path towards a better understanding of the properties of the QCD medium.}

\FullConference{%
  HardProbes2020\\
  1-6 June 2020\\
  Austin, Texas}

\begin{document}
\maketitle

\section{Introduction and overview}
Ultra-relativistic heavy-ion collisions (URHIC) produce a novel state of matter composed of deconfined quarks and gluons known as the quark gluon plasma (QGP). Characterizing the properties of the QGP is one of the main goals for modern URHIC experiments, such as those currently running at the Relativistic Heavy-Ion Collider (RHIC) and the Large Hadron Collider (LHC). Electroweak (EW) probes are key in this regard as they can separate initial state information, following the onset of a heavy-ion collision, from later dynamical evolution of the QGP, given how small EW coupling is compared to strong coupling (at energy scales probed by RHIC or the LHC). Given their large masses, weak probes, composed of $W^\pm$ and $Z^0$ gauge bosons, are preferentially produced near the onset of a heavy-ion collision, and thus are sensitive to the modifications of the parton distribution functions (PDFs) inside the nucleus (see e.g. \cite{Acharya:2017wpf} for a recent results about $Z^0$ production). Compared to  $W^\pm$ and $Z^0$,  electromagnetic (EM) probes are typically much lighter (be it real photons or virtual photons decaying into leptons pairs, i.e. dileptons), and are not only emitted at the early stages of the collision,\footnote{Prompt photons and Drell-Yan (DY) dileptons are most directly sensitive to nuclear modifications to the PDFs.} but also throughout the entire evolution of the QGP. Furthermore, as soon as EM probes are created in the QGP, they escape it with negligible rescattering, thus giving precise information about that state of the QGP at their production point. Thus, EM radiation has been considered as a ``clean'' probe of the QGP. As these proceeding are focusing on the properties of the QCD medium, discussion about EM probes will take center stage. 

Among electromagnetic probes, i.e. photons and dileptons, the latter have additional degrees of freedom, the center of mass energy of the lepton pair, or the invariant mass ($M$) of the virtual photon, which allows for a separation between different sources of radiation from the QCD medium. In the low invariant mass region ($M\lesssim 1$ GeV), dilepton radiation is dominated by the hadronic medium (HM), notably from in-medium decays of vector mesons (i.e. $\rho$, $\omega$, and $\phi$), while at intermediate invariant masses ($1\lesssim M\lesssim 2.5$ GeV) radiation stemming directly from the QGP is an important source. The feature-rich spectral shape coming from hadronic resonance contributions to the electromagnetic spectral function \cite{Vujanovic:2019yih}, seen in dilepton invariant mass spectra, facilitates to distinguish HM from the QGP sources, whose spectral shape is smooth with respect to $M$.\footnote{The transverse momentum ($p_T$) spectrum of EM probes is rather featureless, thus it is difficult to distinguish between the HM and QGP sources using the $p_T$ spectrum alone.} Another way to detect radiation directly from the QCD medium, compared to dilepton sources that exit in p-p collisions, is via a sizable dilepton $v_2$, which has not been observed in p-p collisions. Measuring dilepton $v_2$ at intermediate invariant masses would confirm that the high temperature medium expands anisotropically, which can be used to narrow down the speed of sound in that region (see Ref. \cite{Pratt:2015zsa}), provided experimentally ``confounding'' sources are removed. 

Dilepton radiation in $1\lesssim M \lesssim 2.5$ GeV measured experimentally is not solely coming from the QGP, but also includes semi-leptonic decays of open heavy flavors as a ``confounding'' source. The semi-leptonic decay of open heavy (anti)flavor pairs can be precisely studied via a Heavy Flavor Tracker, such at the one installed by the STAR experiment at RHIC. Understanding the contribution from open heavy flavor to dilepton production in that $M$ region is interesting in and of itself, as will be highlighted below. However, distinguishing  between dileptons coming from semi-leptonic decays of open heavy (anti)flavor pairs and those from the QGP allows to study the latter {\it directly}, especially through its $v_2$ at intermediate invariant masses \cite{Vujanovic:2017psb}.\footnote{Note that heavy flavor is not produced within the QGP in practice, given the temperatures accessible to RHIC and LHC collisions.} There are sources to dilepton production coming from decays of heavy quarkonia contributing to the range $M\gtrsim 2.5$ GeV, however such sources will not be the focus here. 

As far as direct photons are concerned, beyond prompt photons, there are photons before and during the hydrodynamical evolution of the QCD medium contributing to the direct photon spectrum \cite{Gale:2020xlg}. 

The last sources of photon and dilepton production stem from late hadronic interaction happening following the hydrodynamical evolution. For direct photons, late decays of hadrons are removed; for dileptons, whether that stage is removed or not depends on the experimental capabilities.    

\section{Electromagnetic radiation from the QCD medium}

The production rate of EM probes, in thermal equilibrium, is given by:
\begin{eqnarray}
\frac{d^4 R_{\ell^+\ell^-}}{d^4 q} &=& -\frac{\alpha^2_{EM}}{\pi^3 M^2} \frac{{\rm Im}\left[\Pi_{EM}(M,{\bf q};T,\mu_i)\right]}{e^{ p\cdot u/T} -1}\nonumber\\
q^0 \frac{d^3 R_\gamma}{d^3 {\bf q}} &=&  -\frac{\alpha_{EM}}{\pi^2} \frac{{\rm Im}\left[\Pi_{EM}(M=0,{\bf q};T,\mu_i)\right]}{e^{p\cdot u/T} -1},
\label{eq:th_rates}
\end{eqnarray}
where $M^2=\left(q^0\right)^2-\vert{\bf q}\vert^2$, $T$ is the temperature, and ${\rm Im}\left[\Pi_{EM}\right]={\rm Im}\left[g_{\mu\nu} \Pi^{\mu\nu}_{EM}\right]$ is the EM spectral function. A dedicated effort has been invested in recent years towards studying the in-equilibrium electromagnetic spectral function using both perturbative and non-perturbative techniques. In parallel to these developments, an equally important endeavor was the inclusion of non-equilibrium effects, through viscous corrections $(\delta R)$ added to the thermal rates in Eq. (\ref{eq:th_rates}), accounting for non-equilibrium deviations owing to, for example, shear and bulk viscosity of the QCD medium, in both partonic and hadronic sectors.

\subsection{In-equilibrium electromagnetic rates}
Using perturbative QCD (pQCD), the electromagnetic spectral function has been calculated at next-to-leading order (NLO) for both photons \cite{Ghiglieri:2013gia,Ghiglieri:2016tvj} and dileptons \cite{Laine:2013vma,Ghisoiu:2014mha,Ghiglieri:2014kma}. These efforts ultimately culminated in the results found in Refs. \cite{Jackson:2019mop,Jackson:2019yao} where using the NLO pQCD EM spectral function was devised combining results both above and below the light cone, as well as employing longitudinal and transverse channels with respect to spatial momentum. Using these new developments, Ref. \cite{Jackson:2019yao} shows an unprecedented agreement between NLO pQCD EM spectral function and lattice QCD results of Ref. \cite{Brandt:2017vgl,Brandt:2019shg}, for both quenched and unquenched lattice calculations; thus ushering a new era of precision calculations of EM spectral functions. 

In the hadronic sector, many-body effective Lagrangians are used to describe hadronic interactions in the QCD medium, while the inclusion of the EM interaction is achieved through the coupling to vector mesons, well described by the vector meson dominance (VDM). The leading contribution to the EM spectral function stems from tree-level scattering matrix elements, which can also be used in a Boltzmann description of the hadronic scattering rates. More specifically, the mesonic contribution to photon production using SU(3) Massive Yang-Mills theory was discussed in \cite{Turbide:2003si}, while tree-level scattering-based approach for dileptons in a medium composed of pions and nucleons, is discussed in \cite{Eletsky:2001bb,Martell:2004gt,Vujanovic:2009wr}. In addition to tree-level hadronic interactions, higher order correction to the EM spectral function have also been included \cite{Rapp:1999ej} (see also \cite{Rapp:2009yu} for a more recent review). The baryonic contribution to the photon production rates are also included in the EM spectral function of Refs. \cite{Rapp:1999ej,Rapp:2009yu}, while parametrizations of photon production rates from both sources can be found in Refs. \cite{Turbide:2003si, Heffernan:2014mla}. On the dilepton side, additional  baryonic interactions included in Refs. \cite{Rapp:1999ej,Rapp:2009yu} compared to \cite{Eletsky:2001bb,Martell:2004gt,Vujanovic:2009wr} were found to be important contributors to vector meson spectral functions. These findings have recently been extended to the chiral partner of the $\rho$ meson ---  the $a_1$ --- thus opening the possibility to study the axial vector spectral function of the $a_1$. As the ($\rho,a_1$) constitutes a chiral partner pair, a study of the vector-axial vector spectral function \cite{Hohler:2013eba} as temperature increases shows that the $\rho$ and $a_1$ start to overlap at temperature $T=170$ MeV. This is one of the first signs of chiral symmetry restoration appearing in hadronic Lagrangians. Another promising approach to studying chiral symmetry restoration relies on the non-perturbative functional renormalization group (FRG) \cite{Tripolt:2017zgc}. Using FRG, it was found that the $\rho$ and $a_1$ spectral function overlap at $T=300$ MeV. However, the calculation of Ref. \cite{Tripolt:2017zgc} is lacking baryonic degrees of freedom, which may bring the overlap temperature closer to the pseudo-critical temperature $T=156 \pm 1.5$ MeV \cite{Bazavov:2018mes}.     

\subsection{Dissipative corrections to electromagnetic rates}
Modern hydrodynamical modeling of the QCD medium involves dissipative phenomena such as viscosity, thus requiring modifications to the equilibrated EM rates. Introducing dissipative degrees of freedom to electromagnetic production has been done for all tree-level matrix elements through their Boltzmann kinetic theory representation. Indeed, under the Boltzmann approximation, the thermal photon rates are:  
\begin{eqnarray}
\frac{d^3 R_\gamma}{d^3 {\bf q}} = \int \frac{d^3 p_1}{2p^0_1 (2\pi)^3} \frac{d^3 p_2}{2p^0_2 (2\pi)^3}  \frac{d^3 p_3}{2p^0_3 (2\pi)^3} (2\pi)^4 \delta^{(4)}(p_1+p_2-p_3-q) f_{\bf p_1} f_{\bf p_2} \frac{\left\vert \mathcal{M} \right\vert^2}{2q^0(2\pi)^3} (1\pm f_{\bf p_3}),
\end{eqnarray}
where $f_{\bf p_i}$ are the equilibrium distributions of scattering particles. To obtain dissipative corrections, Refs. \cite{Dion:2011pp,Shen:2014nfa,Paquet:2015lta} use the Chapman-Enskog/14-moment approximations as ans\" atze, which results into $f\to f+\delta f$. The effects of both bulk and shear viscous corrections are encapsulated in $\delta f$. The viscous correction $\delta R$ to the rate is obtained by expanding to linear order in $\delta f$ to the deviation from the equilibrium distribution $f$ \cite{Dion:2011pp,Shen:2014nfa,Paquet:2015lta}. Going beyond the tree-level interactions, in particular including the Landau-Pomeranchuck-Migdal effect, was only done for pQCD photon rates \cite{Hauksson:2017udm}, though the implications for phenomenology of such rates have yet to be explored in full.

As far as dilepton rates are concerned, the effects of the above-mentioned $\delta f$ viscous correction was done in the leading density expansion \cite{Eletsky:2001bb,Martell:2004gt,Vujanovic:2009wr} presented in Refs. \cite{Vujanovic:2013jpa,Vujanovic:2017psb,Vujanovic:2019yih}. 

\section{Selected phenomenology of electromagnetic probes}
\begin{figure}[!h]
\begin{tabular}{cc}
\includegraphics[width=0.5\textwidth]{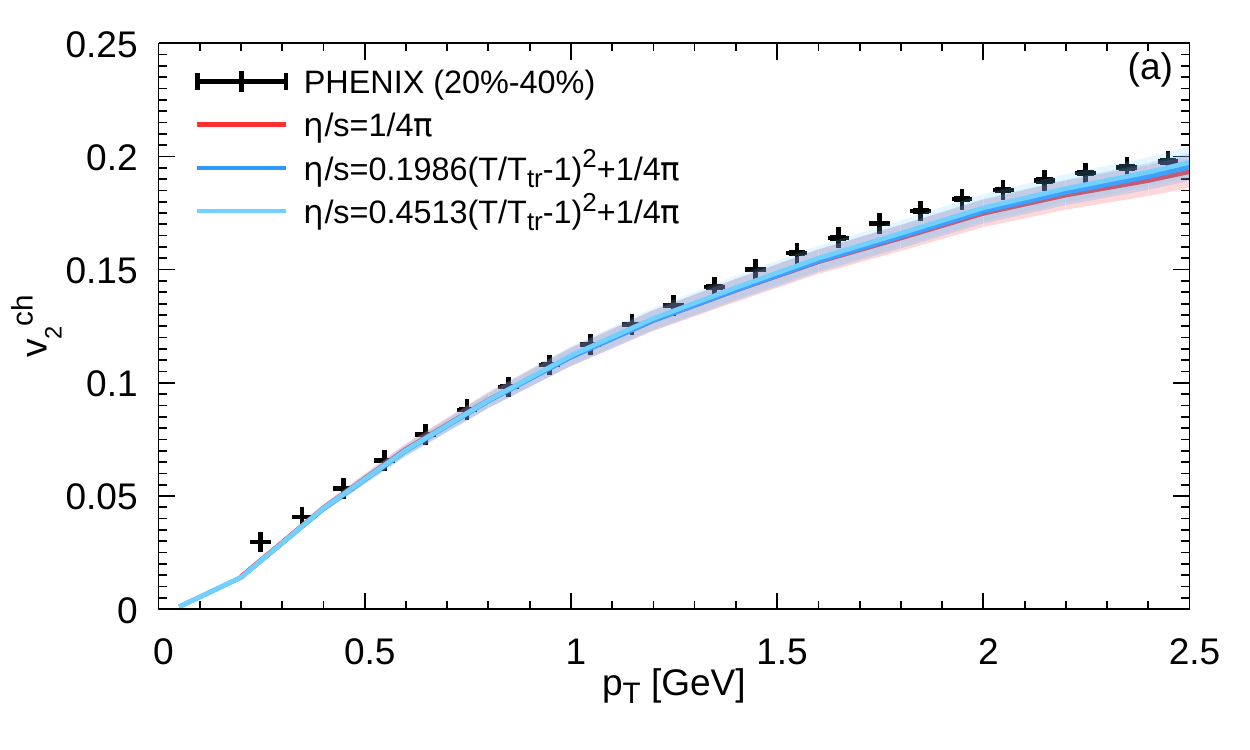} & \includegraphics[width=0.5\textwidth]{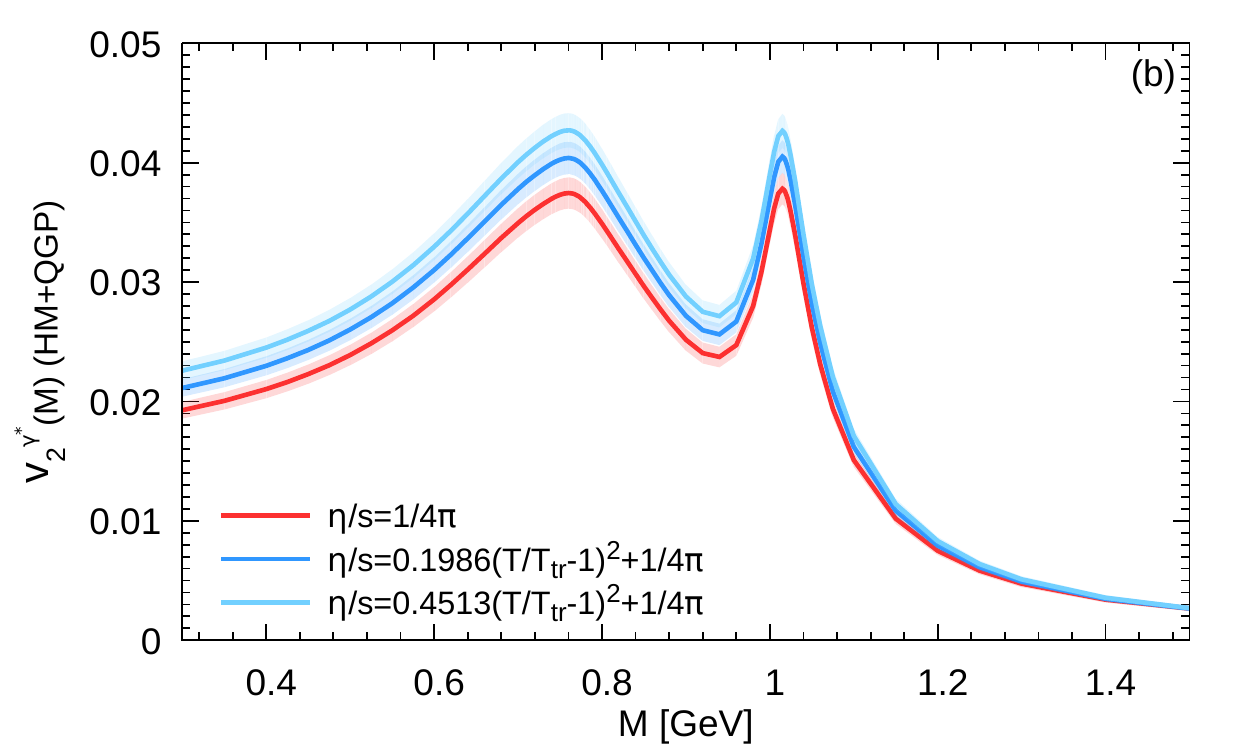} 
\end{tabular}
\caption{Charged hadron $v^{\rm ch}_2$ and dilepton $v_2(M)$ within $20-40$\% centrality at top RHIC collision energy taken from \cite{Vujanovic:2017psb}. This dilepton calculation includes sources from the HM and the QGP. The same underlying hydrodynamical simulations are used for both observables.}
\label{fig:v2_M_vs_ch}
\end{figure}

Except for the NLO pQCD EM rates of Ref. \cite{Jackson:2019yao}, the above-mentioned EM production rates have been used in many dynamical models, and compared against experimental data. At top RHIC or LHC energies, these are ranging from early-time dynamics modeled by e.g. effective kinetic theory \cite{J_F_P_these}, to late-time dynamics modeled by hadronic transport\footnote{At lower beam energies, hadronic transport can and, in fact, is used to describe much of the evolution of the system, leaving little room for hydrodynamics or other effective kinetic theories at those lower energies.} \cite{Endres:2015egk,Endres:2016tkg,Staudenmaier:2017vtq,Schafer:2019edr}, with hydrodynamics used to describe the evolution between these extremes. Alternative dynamical modeling schemes to the one just outlined consist, for instance, Parton-Hadron-String Dynamics (PHSD) \cite{Song:2018xca,E_B_these}, Boltzmann approach to multiparton scatterings (BAMPS) \cite{Greif:2016jeb} and so on. The comparison of such simulations against experimental data is very useful to learn about, and even constrain, various properties of the QCD medium, such as its viscous transport coefficients. Doing so in the hydrodynamical context has shown that the penetrating nature of EM probes allows them to be quite sensitive to transport coefficients. For example, dileptons are can be more sensitive to the temperature-dependent specific shear viscosity $\frac{\eta}{s}(T)$ than hadrons are. This is displayed in Fig. \ref{fig:v2_M_vs_ch} using a quadratic form of $\frac{\eta}{s}(T)$ (for details see Ref. \cite{Vujanovic:2017psb}). 
This result is particularly important in light of the recent Bayesian analysis \cite{J_F_P_QM19} where $\frac{\eta}{s}(T)$ is poorly constrained at high temperatures. For this sensitivity of dileptons to be fully exploited however, accurate measurements of dilepton $v_2$, such as those being planned in Ref. \cite{Citron:2018lsq}, are needed. Electromagnetic probes are also sensitive to other viscous transport coefficients, such as shear relaxation time \cite{Vujanovic:2016anq} or bulk viscosity \cite{Paquet:2015lta,Vujanovic:2019yih} of the hydrodynamical medium, thus using EM probes together with hadrons in a Bayesian analysis should be considered in the path towards increased constraints on transport coefficients of QCD media.

There are other sources of EM probes, beyond those produced hydrodynamically, that need to be included as well. For purposes of extracting properties of the QGP, an important source of photons emission stems from pre-hydrodynamical simulations. On the dilepton side, the semi-leptonic decays of open heavy (anti-)flavor pairs contribute for $1\lesssim M\lesssim 2.5$ GeV \cite{Vujanovic:2013jpa}. 

As the heavy (anti-)quark pair traverses the QGP, the amount of energy/momentum exchange depends both kinematics (such as the virtuality of the heavy quark), as well as on the local QGP properties, such as $\hat{q}$ or $\eta_D$.\footnote{$\hat{q}$ measures the amount of momentum diffusion, while $\eta_D$ measures instead the amount of momentum drag.} Describing the evolution of the heavy quark within the QGP is best achieved using an agnostic framework that allows for different models to be used (and also compared) depending on the kinematic regime. This is currently being pursued inside the JETSCAPE Collaboration \cite{Vujanovic:2020wuk,Fan_these}. Following hadronization, the dilepton signal originates from the decay of the open heavy (anti-)flavor {\it pair} and is thus different from other semi-leptonic decay measurements that do not consider open heavy flavors as a pair. The heavy quark energy/momentum interaction with the QGP also generates dilepton $v_2$ \cite{Vujanovic:2013jpa}. Thus, measuring dilepton $v_2$ in the intermediate mass region is crucial as it gives novel information about $\hat{q}$ or $\eta_D$ from open heavy flavor decay pairs, while direct radiation from the QGP gives better access to bulk properties such as viscosities.     

\begin{figure}[!h]
\begin{tabular}{cc}
\includegraphics[width=0.5\textwidth]{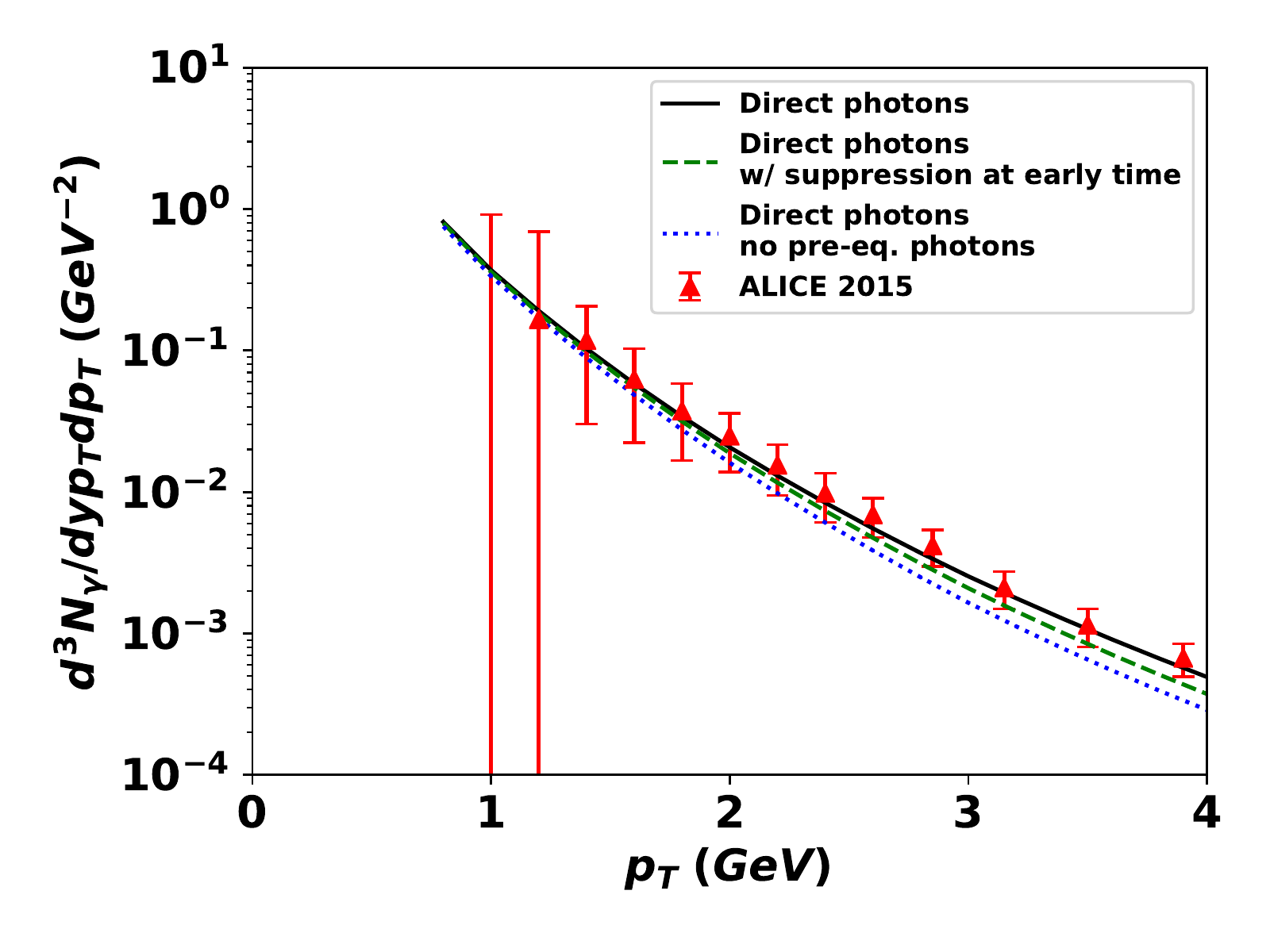} & \includegraphics[width=0.5\textwidth]{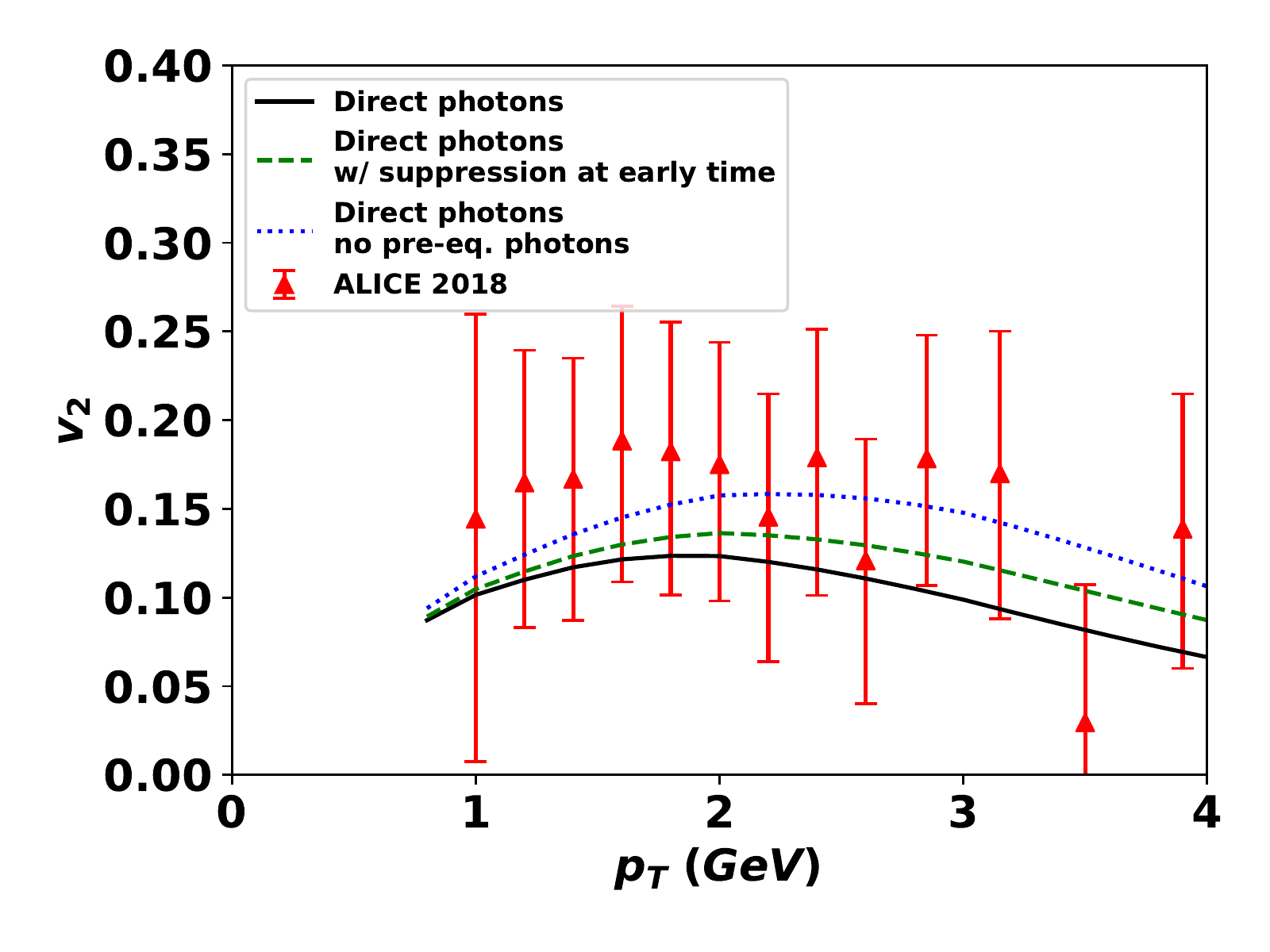} 
\end{tabular}
\caption{Direct photon yield and $v_2$ within $20-40$\% centrality for $\sqrt{s_{NN}}=2.76$ TeV Pb-Pb collisions carried out at the LHC \cite{Gale:2020xlg}.}
\label{fig:photon_yield_v2}
\end{figure}
Starting from studies on prompt and hydrodynamical photons \cite{Paquet:2015lta}, novel studies include photon production from a pre-hydrodynamical evolution \cite{Gale:2020xlg,J_F_P_these}. These pre-hydordynamical simulations not only provide a non-trivial initial condition for hydrodynamics, but also possess important physics processes for photon production, notably dynamical quark generation. Indeed, a dynamical quark production bridges the gap between the gluon-dominated PDF sampled/created at the onset of high energy collisions, and the hydrodynamically evolved QGP, that assumes thermally equilibrated quarks. From that perspective, K\o MP\o ST is an interesting effective kinetic theory as it provides a mechanism to dynamically generate quarks.  Figure  \ref{fig:photon_yield_v2} is showing the sensitivity of direct photon production to the EM radiation before hydrodynamics, and thus to the dynamically generated quarks. The effects of photon production during  K\o MP\o ST in Fig.~\ref{fig:photon_yield_v2} are strong enough to affect total direct photon spectra and $v_2$. This is a promising avenue to explore further in the future. 

\section{Conclusion and outlook}
As all of the major ingredients regarding photon/dilepton production calculations are largely in place, more focus should now be directed towards combining these different calculations together, while gearing towards a comprehensive/simultaneous understanding of EM probes and hadronic observables. Of course, theoretical improvements pertaining to the production rates (via e.g. viscous correction prescription used), or enhancements in the dynamical modeling (e.g. pre-hydrodynamical evolution/EM production) should happen in parallel. Those theoretical advancements are useful both in improving the understanding of EM production, and serve as a source of theoretical uncertainty to the already available sources to be used within a future Bayesian analysis. A combined calculation within a Bayesian model-to-data comparison will not only allow for a better constraint on transport coefficients, such as $\hat{q}$ or $\frac{\eta}{s}(T)$, but also be employed to give more credence to certain models. For instance, one can use Bayesian model selection on a dilepton calculation that contain chiral symmetry restoration effects compared to the one without, to ascertain whether experimental data favors either calculation. Bayesian analyses could also help devise whether better measurements are needed in certain areas (see e.g. Ref. \cite{Sangaline:2015isa}). A precision measurement of dilepton $v_2$ is thus crucially needed, such as that being planned in Ref. \cite{Citron:2018lsq}.

{\bf Acknowledgements}: This work was supported by the Natural Sciences and Engineering Research Council of Canada, and by the National Science Foundation (in the framework of the JETSCAPE Collaboration) through award No. ACI-1550300.

\bibliographystyle{h-physrev3.bst}
\bibliography{references}

\end{document}